\documentclass[preprint,prc,showpacs]{revtex4-1}
\usepackage{amsfonts}

\begin{document}

\title{Relativistic Coulomb scattering of spinless bosons}
\author{M. G. Garcia}
\email{marcelogarcia82@gmail.com}
\author{A. S. de Castro}
\email{castro@pq.cnpq.br}
\date{11 February 2015}

\begin{abstract}
The relativistic scattering of spin-0 bosons by spherically symmetric
Coulomb fields is analyzed in detail with an arbitrary mixing of vector and
scalar couplings. It is shown that the partial wave series reduces the
scattering amplitude to the closed Rutherford formula exactly when the
vector and scalar potentials have the same magnitude, and as an
approximation for weak fields. The behavior of the scattering amplitude near
the conditions that furnish its closed form is also discussed. Strong
suppressions of the scattering amplitude when the vector and scalar
potentials have the same magnitude are observed either for particles or
antiparticles with low incident momentum. We point out that such strong
suppressions might be relevant in the analysis of the scattering of fermions
near the conditions for the spin and pseudospin symmetries. From the complex
poles of the partial scattering amplitude the exact closed form of
bound-state solutions for both particles and antiparticles with different
scenarios for the coupling constants are obtained. Perturbative breaking of
the accidental degeneracy appearing in a pair of special cases is related to
the nonconservation of the Runge-Lenz vector.
\end{abstract}

\maketitle

\affiliation{Universidade Estadual Paulista - Campus de Guaratinguet\'{a},
Departamento de F\'{\i}sica e Qu\'{\i}mica, 12516-410 Guaratinguet\'{a} SP,
Brazil}

\affiliation{Departamento de F\'{\i}sica e Qu\'{\i}mica, Universidade
Estadual Paulista, 12516-410 Guaratinguet\'{a}, S\~ao Paulo, Brazil}

\section{Introduction}

Elastic scattering experiments of a nucleon or electron by atomic nuclei is
of the greatest importance because they can furnish details of the current
and charge distributions within the nucleus by measuring the deviation of
the actual scattering from that due to a point nucleus. The nonrelativistic
quantum solution for the elastic scattering of particles in Coulomb fields
furnishes the classical Rutherford scattering amplitude. This result was
first obtained by Gordon \cite{gor} via separation in parabolic coordinates.
Then, Mott found the partial wave series in terms of phase shifts by using
the Schr\"{o}dinger equation \cite{mot1} as well as the Dirac equation with
the Coulomb field as a time component of a Lorentz vector \cite{mot2}. In
this last case, though, the scattering amplitude is not known in a closed
form.

The elastic scattering of a spinless particle such as a pion or a kaon by a
point nucleus is also of fundamental importance and this problem has
received some attention in the literature. Kang and Brown \cite{kan}
approached the scattering amplitude by using the Klein-Gordon (KG) equation
with the time component of a Coulomb field and found a perturbative
expansion on the coupling constant to the third order whereas Hetherington
\cite{het} analyzed questions regarding the convergence of the partial wave
series, and Rawitscher \cite{raw} examined deviations of the KG from the Schr%
\"{o}dinger scattering for small velocities. Omission of a term involving
the square of the field in the KG equation has been repeated often \cite{jac}%
, and in another publication the contribution of that term only appeared in
the lowest-order terms of the partial wave series \cite{eis}. Latter,
Cooper, Jeppesen and Johnson \cite{coo} filled this gap with an analytic
approximate expression taking into account arbitrarily high angular
momentum. Experimental data from the scattering of kaons by nuclei were
analyzed by Hill, Hetherington and Ravenhall \cite{hil} using the KG
equation with a vector Coulomb field plus a scalar Woods-Saxon field.
Jansen, Pusch and Soff \cite{sof1} investigated the continuum solutions of
the KG equation with vector and scalar couplings for a number of fields,
including the Coulomb field, but their work was not concerned with the
scattering amplitude problem. As a matter of fact, the bound-state solutions
had already received attention before \cite{sof2} (see also Ref. \cite{gre}%
). The mixed vector-scalar Coulomb field was also approached in arbitrary
dimensions with full attention to the bound states \cite{liu}, and later to
the scattering states and their phase shifts \cite{che}. Even so, the
authors of Ref. \cite{che} did not examine in detail the full properties of
the scattering amplitude and for that reason they missed the opportunity to
explore some of its important attributes.

Physical systems with fermions subject to vector and scalar fields are not
uncommon in the literature. The Dirac equation with vector and scalar fields
has been originally used in an attempt to describe the dynamics between a
quark and antiquark in order to calculate meson masses \cite{smi}. Later,
the mixing vector-scalar was used for investigating the implications of a
tiny contribution of the scalar Coulomb potential to atomic spectroscopy
\cite{raf}. The nucleon-nucleus scattering has been described quite well
with strong repulsive vector and attractive scalar fields since 1980s (see
e.g. \cite{ham} and references therein). The Dirac Hamiltonian with vector
and scalar fields with couplings of equal magnitude is invariant under a
SU(2) algebra \cite{bel} and vector and scalar fields with near equality in
the magnitudes have been used as nuclear mean-field potentials \cite{gin}
(related to the pseudospin symmetry), and in the quark model \cite{pag} as
well as for considering the possibility of antifermions in nuclei \cite{zho}
(related to the spin symmetry). Indeed, most studies of spin and pseudospin
symmetries are concerned with bound states, and it seems that the
experimental data on nucleon-nucleus scattering does not exhibit the
features of the approximate pseudospin symmetry \cite{gin}, \cite{bow},\cite%
{lee}. The authors of Ref. \cite{lee} suggest that the long-range nature of
the Coulomb potential is responsible for breaking the pseudospin symmetry.
Nevertheless, the approximate realization of that symmetry has been found by
other authors using different kinds of analysis \cite{lu}. Recently, the
perturbative nature of both symmetries with Coulomb fields were discussed in
Ref. \cite{asc1} regarding bound states. It has also been shown that, for
fields of any shape, the conditions on the coupling constants that originate
the pseudospin and spin symmetries make spin-1/2 and spin-0 particles to
have the same energy spectrum due to the disappearance of the spin-orbit
coupling and the Darwin terms of either the upper component (spin symmetry)
or the lower component (pseudospin symmetry) of the Dirac spinor \cite{asc2}%
. It is the absence of the spin-orbit term for one of the components of the
Dirac spinor that explains the existence of quasi-degenerate pseudospin
doublets in certain nuclei, and the existence of nearly degenerate spin
doublets in heavy-light quark meson.

In applications, the scattering of mesons by a nucleus is described by a
short-range field and an additional vector Coulomb field $V_{v}=\hbar
cZ\alpha /r$ acting outside the region of the nuclear matter ($Z$ the
product of the nuclear and mesonic charges, and $\alpha $ the fine-structure
constant). Addition of a short-range scalar field, related to the exchange
of mediating massive bosons, is important not only for scattering states but
also for describing the spectrum of a meson immersed in a nuclear
environment.

The purpose of this work is to consider the partial wave analysis for the
relativistic elastic scattering of spin-0 bosons by a Coulomb field in the
presence of mixing of vector and scalar couplings. The unified analysis of
the KG equation with vector and scalar Coulomb fields allows one to find the
conditions under which closed forms for the scattering amplitude can be
found that apply to both scattering states and bound states. The addition of
a scalar coupling to the usual vector coupling increases its interest
because the partial wave series can be exactly summed when the vector and
scalar couplings have the same magnitude. The Rutherford formula can also be
retrieved as an approximation for very small coupling constants, as it
happens in the nonrelativistic limit of the theory. The solution of the KG
equation is expressed in terms of the Whittaker functions, and the
scattering amplitude for small pure vector and pure scalar couplings, as
well as the small deviation from the exact formula for vector and scalar
couplings with the same magnitude, is calculated in a perturbative way. The
exact bound-state solutions for a restrict range of coupling constants are
obtained from the complex poles of the partial scattering amplitude. For
such bound states the eigenenergies are expressed in terms of solutions of a
second-degree algebraic equation and the eigenfunctions are expressed in
terms of the generalized Laguerre polynomials. For both kinds of stationary
states we present a detailed study of some interesting particular cases.
Surprisingly, there are strong suppressions of the scattering amplitude when
the vector and scalar fields have equal magnitude either for particles or
antiparticles with low incident momentum. Furthermore, we show that the
accidental degeneracy for bound states appearing in the nonrelativistic
limit and when the vector and scalar couplings have the same magnitude,
related to the conservation of the Runge-Lenz vector, is broken
perturbatively. Finally, we suggest that the strong suppressions mentioned
above might be relevant in the analysis of the scattering of fermions and
antifermions near the conditions required for the spin and pseudospin
symmetries.

\section{Vector and scalar couplings in the Klein-Gordon equation}

The time-independent KG equation for a spinless particle with mass $m$ and
energy $E$ under the influence of external vector ($V_{v}$) and scalar ($%
V_{s}$) fields reads%
\begin{equation}
\left[ \left( \hbar c\right) ^{2}\nabla ^{2}+\left( E-V_{v}\right)
^{2}-\left( mc^{2}+V_{s}\right) ^{2}\right] \phi =0  \label{KG}
\end{equation}%
Notice that the scalar field is coupled to the mass in accordance with the
substitution $m\rightarrow m+V_{s}/c^{2}$. This prescription allows the
analysis of repulsive as well attractive scalar couplings in the same
framework and furnishes the proper nonrelativistic limit of the KG equation
(fields small compared to $mc^{2}$ and $E\simeq mc^{2}$, and an effective
field $V_{v}+V_{s}$), in contrast with the rule $m^{2}\rightarrow
m^{2}+V_{s}^{2}/c^{4}$ employed in \cite{sof1}-\cite{gre}. If one considers
spherically symmetric fields then $\phi \left( \overrightarrow{r}\right) $
can be factorized as%
\begin{equation}
\phi _{\nu lm_{l}}\left( \overrightarrow{r}\right) =\frac{u_{\nu }\left(
r\right) }{r}Y_{lm_{l}}\left( \theta ,\varphi \right)  \label{fact}
\end{equation}%
where $Y_{lm_{l}}\left( \theta ,\varphi \right) $ is the usual spherical
harmonic, $\nu $ denotes the principal quantum number plus other possible
quantum numbers, and $u_{\nu }\left( r\right) $ obeys the radial equation%
\[
\frac{d^{2}u_{\nu }}{dr^{2}}+\left[ \frac{V_{v}^{2}-V_{s}^{2}-2\left(
EV_{v}+mc^{2}V_{s}\right) }{\left( \hbar c\right) ^{2}}\right.
\]%
\begin{equation}
\qquad \qquad \qquad \qquad \left. -\frac{l\left( l+1\right) }{r^{2}}+k^{2}%
\right] u_{\nu }=0
\end{equation}%
in which $\hbar ck=\sqrt{E^{2}-m^{2}c^{4}}$. It may well be worthwhile to
note that the spectrum changes its sign if $V_{v}$ does, and that it is
symmetrical about $E=0$ if $V_{v}=0$. Notice also that the spectrum presents
a degeneracy of order $2l+1$ with respect to $m_{l}$ due to the spherical
symmetry of the fields. Because $\nabla ^{2}\left( 1/r\right) =-4\pi \delta
\left( \overrightarrow{r}\right) $, unless the fields contain a delta
function at the origin, one must impose the homogeneous Dirichlet condition $%
u_{\nu }(0)=0$ \cite{bay}. On the other hand, if both fields vanish at large
distances the solution $u_{\nu }$ has the asymptotic behavior $e^{ikr}$ as $%
r\rightarrow \infty $. Therefore, scattering states only occur if $%
|E|>mc^{2} $ ($k\in
%TCIMACRO{\U{211d} }%
%BeginExpansion
\mathbb{R}
%EndExpansion
$) whereas bound states might occur only if $|E|<mc^{2}$ ($k=i|k|$).

\section{Coulomb fields}

Antecipating possible future physics applications, a set of mathematical
conditions will be explored below. When the vector and scalar fields are of
Coulomb type, i.e., they are of the form $V_{v}=\hbar cg_{v}/r$ and $%
V_{s}=\hbar cg_{s}/r$, the use of the abbreviations%
\begin{equation}
\gamma _{_{l}}=\sqrt{\left( l+1/2\right) ^{2}+g_{s}^{2}-g_{v}^{2}},\quad
\quad \eta =\frac{Eg_{v}+mc^{2}g_{s}}{\hbar ck}
\end{equation}%
and the change $\zeta =-2ikr$ allow to write the radial KG equation in form
of the Whittaker equation
\begin{equation}
\frac{d^{2}u_{\nu }}{d\zeta ^{2}}+\left( -\frac{1}{4}-\frac{i\eta }{\zeta }+%
\frac{1/4-\gamma _{_{l}}^{2}}{\zeta ^{2}}\right) u_{\nu }=0
\end{equation}%
with linearly independent solutions $M_{-i\eta ,\gamma _{_{l}}}\left( \zeta
\right) $ and $W_{-i\eta ,\gamma _{_{l}}}\left( \zeta \right) $ behaving
like $\zeta ^{1/2+\gamma _{_{l}}}$ and $\zeta ^{1/2-\gamma _{_{l}}}$ close
to the origin, respectively \cite{abr}. Because $u_{\nu }(0)=0$, one has to
consider the solution $u_{\nu }$ proportional to \cite{abr}
\begin{equation}
M_{-i\eta ,\gamma _{_{l}}}\left( \zeta \right) =e^{-\zeta /2}\zeta
^{1/2+\gamma _{_{l}}}M(1/2+\gamma _{_{l}}+i\eta ,1+2\gamma _{_{l}},\zeta )
\end{equation}%
Here,%
\begin{equation}
|g_{v}|<\sqrt{1/4+g_{s}^{2}}  \label{c1}
\end{equation}%
and $M(a,b,z)$ is the confluent hypergeometric function%
\begin{equation}
M(a,b,z)=\sum\limits_{j=0}^{\infty }\frac{\Gamma \left( a+j\right) \Gamma
\left( b\right) }{\Gamma \left( b+j\right) \Gamma \left( a\right) }\frac{%
z^{j}}{j!}
\end{equation}%
Using the asymptotic behavior of $M(a,b,z)$ for large $|z|$ and $-\frac{3\pi
}{2}<\arg z\leq -\frac{\pi }{2}$ \cite{abr}

\begin{equation}
M(a,b,z)\simeq \frac{\Gamma \left( b\right) }{\Gamma \left( b-a\right) }%
e^{-i\pi a}z^{-a}+\frac{\Gamma \left( b\right) }{\Gamma \left( a\right) }%
e^{z}z^{a-b}  \label{asy}
\end{equation}%
one can show that for very large $r$ ($|k|r>>1$)
\begin{equation}
u_{\nu }\left( r\right) \simeq \sin (kr-l\pi /2+\delta _{l})
\end{equation}%
with the relativistic Coulomb phase shift $\delta _{l}=\delta _{l}\left(
\eta \right) $ given by%
\begin{equation}
\delta _{l}=\frac{\pi }{2}\left( l+1/2-\gamma _{_{l}}\right) +\arg \Gamma
\left( 1/2+\gamma _{_{l}}+i\eta \right)  \label{ps}
\end{equation}

It is instructive to note that the term $V_{v}^{2}-V_{s}^{2}$ in the KG
equation (\ref{KG}) gives rise to an attractive (repulsive) short-range $%
r^{-2}$ term if one uses the Coulomb field with $|V_{v}|<|V_{s}|$ ($%
|V_{v}|>|V_{s}|$). Notice also that this short-range term does not exist if
the vector and scalar fields have equal magnitude and loses its importance
in the limit of small coupling constants. Only in the absence of the
short-range term the Runge-Lenz vector is a constant of motion \cite{yos}.

\subsection{Scattering states}

For scattering states, the solution of the KG equation (\ref{KG}) has the
asymptotic form%
\begin{equation}
\phi \left( \overrightarrow{r}\right) \simeq e^{ikr\cos \theta }+f(\theta
,\varphi )\frac{e^{ikr}}{r}  \label{amp}
\end{equation}%
where the first term represents a plane wave moving along the direction $%
\theta =0$ toward the scatterer, and the second represents a radially
outgoing wave. For spherically symmetric scatterers, both terms exhibit
cylindrical symmetry about the direction of incidence in such a way that $%
\phi $ and $f$ are independent of $\varphi $. The connection between the
forms (\ref{fact}) and (\ref{amp}) allows us to write the scattering
amplitude as a partial wave series%
\begin{equation}
f\left( \theta \right) =\sum\limits_{l=0}^{\infty }\left( 2l+1\right)
f_{l}P_{l}\left( \cos \theta \right)  \label{f}
\end{equation}%
where $P_{l}$ is the Legendre polynomial of order $l$ and the partial
scattering amplitude is $f_{l}=\left( e^{2i\delta _{l}}-1\right) /\left(
2ik\right) $. With the phase shift (\ref{ps}), up to a logarithmic phase
inherent to the Coulomb field, one finds%
\begin{equation}
2ikf_{l}=-1+e^{i\pi \left( l+1/2-\gamma_{ _{l}}\right) }\frac{\Gamma \left(
1/2+\gamma_{ _{l}}+i\eta \right) }{\Gamma \left( 1/2+\gamma_{ _{l}}-i\eta
\right) }  \label{phase}
\end{equation}%
The series (\ref{f}) can be summed when $\gamma_{ _{l}}=l+1/2$ the closed
form being \cite{mot1} (see also \cite{lin})%
\begin{equation}
f\left( \theta \right) =-\eta \frac{\Gamma \left( 1+i\eta \right) }{\Gamma
\left( 1-i\eta \right) }\frac{\exp \left( -i\eta \ln \sin ^{2}\theta
/2\right) }{2k\sin ^{2}\theta /2},\quad \theta \neq 0  \label{AMP}
\end{equation}%
which gives the well-known Rutherford scattering formula for the
differential cross section in classical and nonrelativistic quantum
mechanics. It is worthwhile to note that this last equation is exact if one
considers $|g_{v}|=|g_{s}|$. Also, it is appropriate as an approximation for
small coupling constants.

In order to study the behavior of the scattering amplitude near the
conditions that furnish its closed form, we will perform the sum\ (\ref{f})
for small $g_{v}$ and $g_{s}$, convenient for exploring the nonrelativistic
limit. The expansion near the condition $|g_{v}|=|g_{s}|$ is made easier if
one defines $g_{\Delta }$ and $g_{\Sigma }$ by
\begin{equation}
g_{\Delta }=g_{v}-g_{s},\quad g_{\Sigma }=g_{v}+g_{s}  \label{def}
\end{equation}%
The expansions of the sum (\ref{f}) out to next-to-leading order can be
carried out by using a pair of properties of the gamma function \cite{abr}: $%
\Gamma \left( 1+z\right) =z\Gamma \left( z\right) $ and $\ 1/\Gamma \left(
z\right) \simeq ze^{\gamma _{_{\!e}}z}$ for $z\simeq 0$, where $\gamma
_{_{\!e}}$ is Euler's constant. In addition, we use a few mathematical
identities taken from Ref. \cite{kan}, viz.%
\begin{widetext}
\begin{eqnarray}
\sum\limits_{l=0}^{\infty }P_{l}\left( \cos \theta \right) &=&\frac{1}{2\sin
\theta /2}\nonumber \\
&&  \nonumber \\
\sum\limits_{l=0}^{\infty }\left( 2l+1\right) P_{l}\left( \cos \theta
\right) &=&0,\quad \theta \neq 0  \nonumber \\
&&  \nonumber \\
\sum\limits_{l=0}^{\infty }\left( 2l+1\right) P_{l}\left( \cos \theta
\right) \psi \left( l+1\right) &=&-\frac{1}{2\sin ^{2}\theta /2} \\
&&  \nonumber \\
\sum\limits_{l=0}^{\infty }\left( 2l+1\right) P_{l}\left( \cos \theta
\right) \psi ^{2}\left( l+1\right) &=&\frac{\ln \sin \theta /2+\gamma_{_{\!e}} }{\sin
^{2}\theta /2}  \nonumber \\
&&  \nonumber \\
\sum\limits_{l=0}^{\infty }\left( 2l+1\right) P_{l}\left( \cos \theta
\right) \left[ \psi ^{\left( 2\right) }\left( l+1\right) +4\psi
^{3}\left( l+1\right) \right] &=&-6\left( \frac{\ln \sin \theta /2+\gamma_{_{\!e}} }{%
\sin \theta /2}\right) ^{2}  \nonumber
\end{eqnarray}
\end{widetext}where $\psi \left( z\right) =d\ln \Gamma \left( z\right) /dz$
is the digamma (psi) function and $\psi ^{\left( 2\right) }\left( z\right)
=d^{2}\psi \left( z\right) /dz^{2}$. With such identities, comparison of the
expansion of (\ref{AMP}) and (\ref{f}) with $f_{l}$ given by (\ref{phase})
furnishes the desired expansions.

The expansions for $|g_{v}|<<1$ and $|g_{s}|<<1$ are given by
\begin{equation}
f\left( \theta \right) =\left\{
\begin{array}{c}
-\frac{g_{v}E}{2\hbar ck^{2}\sin ^{2}\theta /2}e^{-i\frac{2g_{v}E\left( \ln
\sin \theta /2+\gamma _{_{\!e}}\right) }{\hbar ck}}f_{v},\quad g_{s}=0 \\
\\
-\frac{g_{s}mc^{2}}{2\hbar ck^{2}\sin ^{2}\theta /2}e^{-i\frac{%
2g_{s}mc^{2}\left( \ln \sin \theta /2+\gamma _{_{\!e}}\right) }{\hbar ck}%
}f_{s},\quad g_{v}=0%
\end{array}%
\right.
\end{equation}%
with%
\begin{eqnarray}
f_{v} &=&1-\frac{g_{v}\pi \hbar ck\sin \theta /2}{2E}+\mathcal{O}\left(
g_{v}^{2}\right)   \nonumber \\
&& \\
f_{s} &=&1+\frac{g_{s}\pi \hbar ck\sin \theta /2}{2mc^{2}}+\mathcal{O}\left(
g_{s}^{2}\right)   \nonumber
\end{eqnarray}%
The leading terms in both expansions are consistent with the nonrelativistic
limit with $|E|\simeq mc^{2}$ and the next-to-leading order contributes for
changing the angular distribution. Differences between vector and scalar
couplings appear even in the zeroth-order terms. A remarkable difference
between the natures of vector and scalar couplings appears in the
next-to-leading order: the pure scalar coupling always contributes to
increase (decrease) $|f\left( \theta \right) |$ when the the scalar field is
repulsive (attractive), as for the the pure vector coupling, though, $%
|f\left( \theta \right) |$ increases or decreases depending on the sign of $%
g_{v}E$.

Near the condition $|g_{v}|=|g_{s}|$, with $g_{\Delta }$ and $g_{\Sigma }$
defined in (\ref{def}), one obtains the following perturbative approximations%
\begin{equation}
f\left( \theta \right) =\left\{
\begin{array}{c}
-\frac{g_{\Sigma }\left( E+mc^{2}\right) }{4\hbar ck^{2}\sin ^{2}\theta /2}%
e^{-i\frac{g_{\Sigma }\left( E+mc^{2}\right) \left( \ln \sin \theta
/2+\gamma _{_{\!e}}\right) }{\hbar ck}}f_{\Sigma },\quad g_{\Delta }=0 \\
\\
-\frac{g_{\Delta }\left( E-mc^{2}\right) }{4\hbar ck^{2}\sin ^{2}\theta /2}%
e^{-i\frac{g_{\Delta }\left( E-mc^{2}\right) \left( \ln \sin \theta
/2+\gamma _{_{\!e}}\right) }{\hbar ck}}f_{\Delta },\quad g_{\Sigma }=0%
\end{array}%
\right.
\end{equation}%
with%
\begin{eqnarray}
f_{\Sigma } &=&1-\frac{1}{2}\left[ \frac{g_{\Sigma }\left( E+mc^{2}\right)
\left( \ln \sin \theta /2+\gamma _{_{\!e}}\right) }{\hbar ck}\right] ^{2}+%
\mathcal{O}\left( g_{\Sigma }^{3}\right)   \nonumber \\
&& \\
f_{\Delta } &=&1-\frac{1}{2}\left[ \frac{g_{\Delta }\left( E-mc^{2}\right)
\left( \ln \sin \theta /2+\gamma _{_{\!e}}\right) }{\hbar ck}\right] ^{2}+%
\mathcal{O}\left( g_{\Delta }^{3}\right)   \nonumber
\end{eqnarray}%
The leading terms of these last expansions, for $g_{\Delta }=0$ ($g_{\Sigma
}=0$), reveal consistency with the nonrelativistic limit with $E\simeq
+mc^{2}$ ($E\simeq -mc^{2}$) and that the scattering amplitude is strongly
suppressed when $E\simeq -mc^{2}$ ($E\simeq +mc^{2}$) independently of the
magnitudes of $g_{v}$ and $g_{s}$. The next-to-leading order terms distort
the angular distribution and always contribute to decrease $|f\left( \theta
\right) |$.

\subsection{Bound states}

If $k=i|k|$ ($|E|<mc^{2}$), the partial scattering amplitude becomes
infinite when $1/2+\gamma _{_{l}}-$Im\thinspace $\eta =-N$, where $%
N=0,1,2,\ldots $, due to the poles of the gamma function in the numerator of
(\ref{phase}). Because Im\thinspace $\eta >0$, beyond the constraint (\ref%
{c1}) one has $g_{s}<|g_{v}|$. Remembering the asymptotic behavior of the
confluent hypergeometric function given by (\ref{asy}), we see that the
Whittaker function $M_{\text{\textrm{Im\thinspace }}\eta ,\gamma
_{_{l}}}\left( 2|k|r\right) $ goes asymptotically as $e^{-|k|r}$ as $r$
increases because $M(-N,1+2\gamma _{_{l}},2|k|r)$ is proportional to $%
L_{N}^{\left( 2\gamma _{_{l}}\right) }\left( 2|k|r\right) $, the generalized
Laguerre polynomial of order $N$. Thus, the characteristic pair ($E_{\nu
},u_{\nu }$) represents a spatially localized state explicitly expressed as
\begin{equation}
E_{\nu }=mc^{2}\frac{-\frac{g_{v}}{\nu }\frac{g_{s}}{\nu }\pm \left[
1+\left( \frac{g_{v}}{\nu }\right) ^{2}-\left( \frac{g_{s}}{\nu }\right) ^{2}%
\right] ^{1/2}}{1+\left( \frac{g_{v}}{\nu }\right) ^{2}}  \label{E}
\end{equation}%
\begin{equation}
u_{\nu }=A_{\nu }r^{1/2+\gamma _{_{l}}}e^{-|k|r}L_{\nu -1/2-\gamma
_{_{l}}}^{\left( 2\gamma _{_{l}}\right) }\left( 2|k|r\right)
\end{equation}%
where the quantum number $\nu $ satisfies
\begin{equation}
\nu =\text{Im\thinspace }\eta =N+1/2+\gamma _{_{l}}>1/2  \label{NU}
\end{equation}%
and $A_{\nu }$ is a normalization constant. The energy levels enter from the
continuum to the bound-state gap ($|E|<mc^{2}$) coming from the upper
continuum (related to particle states) or from the lower continuum (related
to antiparticle states).

One may check that the nonrelativistic limit is correct. Indeed, if one sets
$|g_{v}|<<1$ and $|g_{s}|<<1$, one gets%
\begin{equation}
\frac{E_{n}}{mc^{2}}\simeq \pm \left[ 1-\frac{\left( g_{s}\pm g_{v}\right)
^{2}}{2n^{2}}\right] ,\quad g_{s}\pm g_{v}<0  \label{25}
\end{equation}%
in which $n$, contained in the denominator of (\ref{25}), is a positive
integer given by $n=N+l+1=1,2,3,\ldots $ with $l\leq n-1$. Beyond the
degeneracy due to the rotational symmetry, a distinguished degeneracy with
respect to $l$ results in a spectrum $n^{2}$-fold degenerate.

For $g_{v}=\varepsilon g_{s}$ ($g_{s}<0$), the expansion of (\ref{E}) until
next-to-leading order in $\varepsilon $ furnishes

\begin{equation}
\frac{E_{\nu }}{mc^{2}}=\pm \left[ 1-\left( g_{s}/\nu \right) ^{2}\right]
^{1/2}-\varepsilon \left( g_{s}/\nu \right) ^{2}+\mathcal{O}\left(
\varepsilon ^{2}\right)
\end{equation}%
\begin{equation}
\nu =N+1/2+\sqrt{\left( l+1/2\right) ^{2}+g_{s}^{2}}+\nu _{\varepsilon }
\end{equation}%
with%
\begin{equation}
\nu _{\varepsilon }=-\varepsilon ^{2}\frac{g_{s}^{2}}{2\sqrt{\left(
l+1/2\right) ^{2}+g_{s}^{2}}}+\mathcal{O}\left( \varepsilon ^{3}\right)
\end{equation}%
in such a way that the spectrum is symmetrical about $E=0$ in the case of a
pure scalar coupling (the zeroth-order term), as expected. The addition of a
small vector coupling increases the upper bound on $|g_{s}|$, breaks the
symmetry of the spectrum and makes the states of particles (antiparticles)
more tightly bound than the states of antiparticles (particles) if the
vector coupling is attractive (repulsive).

On the other hand, for $g_{s}=\varepsilon g_{v}$ one gets
\begin{equation}
\frac{E_{\nu }}{mc^{2}}=-\frac{{\text{sgn}}(g_{v})}{\left[ 1+\left(
g_{v}/\nu \right) ^{2}\right] ^{1/2}}-\varepsilon \frac{\left( g_{v}/\nu
\right) ^{2}}{1+\left( g_{v}/\nu \right) ^{2}}+\mathcal{O}\left( \varepsilon
^{2}\right)
\end{equation}%
\begin{equation}
\nu =N+1/2+\sqrt{\left( l+1/2\right) ^{2}-g_{v}^{2}}+\nu _{\varepsilon }
\end{equation}%
with%
\begin{equation}
\nu _{\varepsilon }=\varepsilon ^{2}\frac{g_{v}^{2}}{2\sqrt{\left(
l+1/2\right) ^{2}-g_{v}^{2}}}+\mathcal{O}\left( \varepsilon ^{3}\right)
\end{equation}%
In this last case, the positive (negative) energy levels are excluded from
the spectrum if the vector field is repulsive (attractive) when $\varepsilon
=0$, and the energies with the least absolute values tend to $-mc^{2}{\text{%
sgn}}(g_{v})/\sqrt{2}$ when $|g_{v}|$ tends to its limit value equal to $1/2$
in such a way that particles (antiparticles) are always associated with
positive (negative) energy levels. The addition of a scalar contaminant
improves the upper bound imposed on $|g_{v}|$ and increases (decreases) the
binding energy if $\varepsilon <0$ ($\varepsilon >0$). The difference
between bosons and fermions is quite remarkable. Whereas fermions dive into
the continuum of negative energies when they are under the influence of a
pure strong and attractive vector Coulomb field ($E=-mc^{2}$ for $%
g_{v}=Z\alpha =-1$) and pair creation has to be considered, the eigenenergy
for bosons takes on a real value only if $g_{v}=Z\alpha <1/4$.

In order to study the spectrum near the condition $|g_{v}|=|g_{s}|$,
necessarily with $g_{s}<0$, we will perform an expansion of (\ref{E}) and (%
\ref{NU}) for small $g_{\Delta }$ and $g_{\Sigma }$ defined in (\ref{def}).
The dependence of $\nu $ on $g_{\Delta }$ and $g_{\Sigma }$ comes through $%
g_{\Delta }g_{\Sigma }$ so that the expansion for $\nu $ is given by%
\begin{equation}
\nu =N+l+1-\frac{g_{\Delta }g_{\Sigma }}{2l+1}+\mathcal{O}\left( g_{\Delta
}^{2}g_{\Sigma }^{2}\right)
\end{equation}%
For small $g_{\Delta }$ one gets

\begin{equation}
\frac{E_{\nu }^{\left( +\right) }}{mc^{2}}=1-\frac{2\left( g_{\Sigma }/2\nu
\right) ^{2}}{1+\left( g_{\Sigma }/2\nu \right) ^{2}}+\varepsilon _{\Delta }
\label{E1}
\end{equation}%
\begin{equation}
\frac{E_{\nu }^{\left( -\right) }}{mc^{2}}=-1+g_{\Delta }^{2}\frac{1}{2\nu
^{2}}+\mathcal{O}\left( g_{\Delta }^{3}\right)  \label{E2}
\end{equation}%
with%
\begin{equation}
\varepsilon _{\Delta }=g_{\Delta }\frac{2\left( g_{\Sigma }/2\nu \right) ^{3}%
}{\nu \left[ 1+\left( g_{\Sigma }/2\nu \right) ^{2}\right] ^{2}}+\mathcal{O}%
\left( g_{\Delta }^{2}\right)
\end{equation}%
whereas for small $g_{\Sigma }$ one has%
\begin{equation}
\frac{E_{\nu }^{\left( +\right) }}{mc^{2}}=1-g_{\Sigma }^{2}\frac{1}{2\nu
^{2}}+\mathcal{O}\left( g_{\Sigma }^{3}\right)  \label{E3}
\end{equation}

\begin{equation}
\frac{E_{\nu }^{\left( -\right) }}{mc^{2}}=-1+\frac{2\left( g_{\Delta }/2\nu
\right) ^{2}}{1+\left( g_{\Delta }/2\nu \right) ^{2}}+\varepsilon _{\Sigma }
\label{E4}
\end{equation}%
with%
\begin{equation}
\varepsilon _{\Sigma }=-g_{\Sigma }\frac{2\left( g_{\Delta }/2\nu \right)
^{3}}{\nu \left[ 1+\left( g_{\Delta }/2\nu \right) ^{2}\right] ^{2}}+%
\mathcal{O}\left( g_{\Sigma }^{2}\right)
\end{equation}%
Here, the accidental degeneracy already observed in the nonrelativistic
limit also comes to the scene in the zeroth-order of the expansions in $%
g_{\Delta }$ for $E_{\nu }^{\left( +\right) }$, and $g_{\Sigma }$ for $%
E_{\nu }^{\left( -\right) }$. In this particular order, there is no upper
bound on the coupling constant and for an attractive (repulsive) vector
field the energy levels emerge from the upper (lower) continuum for small
couplings and tend asymptotically to the lower (upper) continuum for large
couplings. The accidental degeneracy in the nonrelativistic Coulomb problem
is related to the conservation of the Runge-Lenz vector \cite{yos}. The term
$V_{v}^{2}-V_{s}^{2}$ in the KG equation (\ref{KG}), due to its accompanying
$r^{-2}$ term, violates the conservation of the Runge-Lenz vector. $E_{\nu
}^{\left( +\right) }$ in (\ref{E1}) and $E_{\nu }^{\left( -\right) }$ in (%
\ref{E4}) reveal that the breaking of the accidental degeneracy is
perturbative. This can be seen from the fact that $g_{\Delta }$ and $%
g_{\Sigma }$ act as perturbative parameters in the above equations, so that
one can go continuously from a bound state without accidental degeneracy to
a bound state with accidental degeneracy (the nonperturbed state) as these
parameters go to zero. This does not happen for $E_{\nu }^{\left( -\right) }$
in (\ref{E2}) nor for $E_{\nu }^{\left( +\right) }$ in (\ref{E3}) due to the
inexistence of bound states in the zeroth-order. Related to the perturbative
breaking of the accidental degeneracy, a worthwhile investigation using the
Runge-Lenz vector as a perturbative method for the classical scattering in a
perturbed Coulomb field can be found in Ref. \cite{bar}.

\section{Final remarks}

The partial wave analysis for the elastic scattering of spin-0 bosons by a
Coulomb field with a general mixing of vector and scalar couplings was done
in detail. Not only in the nonrelativistic limit of the theory does the
partial-wave series for the scattering amplitude reduce to that one giving
the Rutherford formula but also when the vector and scalar fields have the
same magnitude. We calculated the scattering amplitude for small pure vector
and pure scalar couplings, as well as deviations from the exact formula for
vector and scalar couplings with the same magnitude, in a perturbative way.
The complex poles of the partial scattering amplitude furnished the exact
bound-state solutions. The eigenenergies of such bound states are solutions
of a second-order algebraic equation, and the corresponding eigenfunctions
are expressed in terms of generalized Laguerre polynomials. We presented a
detailed study of some interesting particular cases of stationary states
paying special attention to the differences between vector and scalar
couplings. Furthermore, we show that when the vector and scalar couplings
have the same magnitude the scattering amplitude for bosons (antibosons) is
strongly suppressed when $g_{v}=-g_{s}$ ($g_{v}=+g_{s}$) and $|E|\simeq
mc^{2}$. The analysis of eigenenergies for $g_{v}\simeq -g_{s}$ and $%
g_{v}\simeq +g_{s}$ showed that the accidental degeneracy seen in the cases $%
g_{v}=-g_{s}$ and $g_{v}=+g_{s}$ is broken perturbatively and that this
breaking is related to the nonconservation of the Runge-Lenz vector.

It seems that the partial wave analysis for the elastic scattering of
fermions by a Coulomb field with a general mixing of vector and scalar
couplings deserves special attention with focus on the possible strong
suppression of the scattering amplitude for antiparticles (particles) in the
case of spin (pseudospin) symmetry in the low-momentum limit. The presence
of an attractive scalar field in addition to the vector field allows to
approach mesonic atoms with a very large $Z$ nucleus and this fact rises
hope for approaching the spectroscopy of mesonic atoms with more realistic
fields with the scalar-vector coupling scheme.

\begin{acknowledgments}
This work was supported in part by means of funds provided by CAPES and CNPq. The authors would like to thank the referee for constructive criticism and valuable suggestions.
\end{acknowledgments}

\end{document}